\documentclass[twoside,twocolumn,english,aps,prl]{revtex4}
\usepackage[T1]{fontenc}
\usepackage[latin1]{inputenc}
\usepackage{babel}
\usepackage{graphics}

\makeatletter

\providecommand{\LyX}{L\kern-.1667em\lower.25em\hbox{Y}\kern-.125emX\@}

\newcommand{\ket}[1]{{| #1 \rangle}}

\makeatother
\begin{document}

\newcommand{\then}{\Rightarrow }

\title{Coulomb scattering in a 2D interacting electron gas and production
of EPR pairs }

\author{D. S. Saraga\( ^{1} \), B. L. Altshuler\( ^{2,3} \),
Daniel Loss\( ^{1} \), and  R. M. Westervelt\( ^{4} \) }

\affiliation{{\footnotesize \( ^{1} \)Department of Physics and Astronomy, University
of Basel, Klingelbergstrasse 82, CH-4056 Basel, Switzerland }}

\affiliation{{\footnotesize \( ^{2} \)Physics Department, Princeton University,
Princeton, New Jersey 08544}}

\affiliation{{\footnotesize \( ^{3} \)NEC Research Institute, 4 Independence
Way, Princeton, New Jersey 08540}}

\affiliation{{\footnotesize \( ^{4} \)Division of Engineering and Applied Sciences,
Harvard University, Cambridge, Massachusetts 02138}}

\begin{abstract}
We propose a setup to generate non-local spin-EPR pairs via pair collisions
in a 2D interacting electron gas, based on constructive two-particle
interference in the spin singlet channel at the $\pi/2$ scattering angle.
We calculate the scattering amplitude 
via the Bethe-Salpeter equation in the ladder approximation and small
$r_{s}$ limit, and find that the Fermi sea 
leads to a substantial renormalization of the bare scattering process.
From the scattering length we estimate the 
current of spin-entangled electrons and show that it is within 
experimental reach.

PACS numbers :  73.23.-b, 71.10.Ca,  3.67.Mn,{\footnotesize \vspace{-5mm}}{\footnotesize \par}
\end{abstract}

\date{\today}

\maketitle

In recent years, the spin degree of freedom of electrons has become of central interest in
semiconductor research \cite{Aws02}.
This is particularly so for spin-based quantum information
processing, where the basic resources are EPR pairs such as non-local spin singlets
formed by two electrons that are spatially separated \cite{Bur00}.
A number of recent publications have described ways to produce such spin-correlated 
two-electron states, as well as orbital entanglement  
\cite{Rec01,Oli02,Cos01,Bee03}. 
Here, we propose a new
concept based on a two-particle interference mechanism that is well-known
from elementary scattering theory \cite{Tay72}:
the  cross-section for two electrons in vacuum, 
given in terms of the scattering amplitude $f$ by
\( \lambda _{S/T}(\theta )=|f(\theta )\pm f(\pi 
-\theta )|^{2} \), 
favors
singlet ($+$) over triplet ($-$) states in the outgoing channel
around the scattering angle $\theta=\pi/2$.
Thus, in principle, two-particle scattering processes can be used
to generate EPR pairs. However, in the context of solid state systems,
the question immediately arises
whether this scattering effect remains operational, and moreover 
observable, in the presence of a Fermi sea consisting of many interacting electrons 
such as a typical two-dimensional electron gas (2DEG) formed in GaAs heterostructures. 
In this Letter, we will show  that within Fermi liquid theory the answer  is affirmative. 

\begin{figure}[!tb]
{\centering \resizebox*{1\columnwidth}{!}{\includegraphics{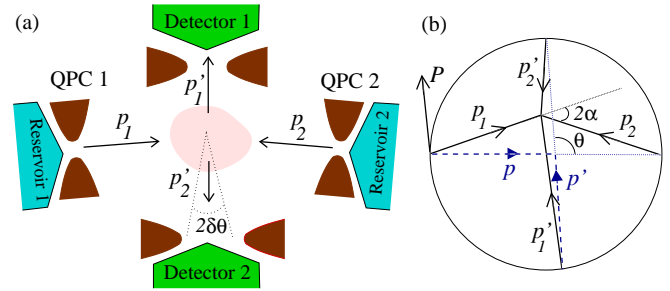}} \par}
\caption{ { \small  \label{figset} 
(a) Proposed setup: two quantum point contacts
(QPC) filter electrons from two reservoirs with initial
momenta \protect\( \mathbf{p}_{1}\simeq -\mathbf{p}_{2}\protect \).
The two detectors (with an aperture angle 
\protect\( 2\delta \theta \protect \)) are placed such that only electrons 
which collide (shaded area)
at a scattering angle around \protect\( \pi /2\protect \)
are registered. Because of interference, the scattering amplitude $f$
vanishes  at \protect\( \pi /2\protect \) for the spin-triplet 
states, allowing only the
spin-entangled singlets to be collected: one electron of the singlet state in detector 1
and its partner in detector 2. The scattering
cross-section and the electron flux could be measured via an AFM tip \cite{Top00}. 
(b) Scattering parameters:
\protect\( \mathbf{P}= \mathbf{p}_{1}+\mathbf{p}_{2}=\mathbf{p}_{1}'+\mathbf{p}_{2}' \) is 
the total momentum, {\small \protect\( \mathbf{p}=(\mathbf{p}_{1}-
\mathbf{p}_{2})/2 \) and \(\mathbf{p}'=(\mathbf{p}_{1}'-\mathbf{p}'_{2})/2\protect \)}
are the relative momenta, and  
\protect\( \theta =\protect \angle (\mathbf{p},\mathbf{p}')\protect \)
 is the scattering angle between them.
The initial (\protect\( \mathbf{p}_{1},\mathbf{p}_{2}\protect \) )
and final  (\protect\( \mathbf{p}_{1}',\mathbf{p}_{2}'\protect \))
momenta are connected by a circle of radius $p=p'$ due to energy and 
momentum conservation.} \vspace{-5mm}
}
\end{figure}

We focus
on 2DEGs since these systems are promising candidates for the 
observation of such effects. Indeed, recent 
experiments \cite{Top00} have demonstrated that in a 2DEG the flow of electrons 
as well as scattering off impurities can be controlled and 
monitored via AFM technology. 
Thus, 
we 
believe that a setup as 
shown in Fig. \ref{figset}(a) is experimentally realizable and should 
allow
the observation of 
the angular and density dependence of the scattering cross section.
Once the EPR pairs are 
created their singlet character can be tested \cite{noisereview}
by a
noise measurement in a beam splitter configuration \cite{Bur00},
or by tests of Bell inequalities \cite{Kaw01}.

We note that the experimental
observation of this type of entanglement would provide  support
for the applicability of Fermi liquid theory to 2D systems, which
has been questioned by Anderson \cite{Anderson}. Indeed, 
one can hardly imagine
singlet pairs of particles that are separated by mesoscopic distances
without well-defined fermionic quasi-particles.

For  electrons incident from unpolarized sources, we expect
the ratio of singlets $\left| S\right\rangle$ to triplets 
$\left| T_{i}\right\rangle$, $i=0, \pm$ to be 1:3, this mixed state 
being described 
by  \( 1/4\left| S\right\rangle 
\left\langle S\right| +1/4\sum _{i=0,\pm }\left| T_{i}\right\rangle 
\left\langle T_{i}\right|  \).
Our goal 
is to calculate the ratio, of scattered 
triplets to scattered singlets, for an aperture angle \( 2\delta \theta  \)
of the detectors around \( \theta =\pi /2 \).
For small \(\delta \theta \)this ratio is
\( R\simeq \delta \theta ^{2}\left| f'/f\right| _{\theta =\pi /2}^{2} \),
where
$f$ is the many-body scattering amplitude and $f'$ its angular derivative. 
Solving the  Bethe-Salpeter equation in the small $r_{s}$ limit
we find that \( \left| f'/f\right| ^{2}\sim 1 \) at \( \theta =\pi /2 \). 
This means that  for 
small 
\( \delta \theta  \)
the scattering process 
produces
dominantly  singlets 
in the direction of the detectors 1 and 2; see Fig. \ref{figset}(a). 
As an experimental check, this ratio can be
increased by
reducing the amount of singlets
in the incoming channels, which in turn can be achieved e.g. by using spin polarized
electron sources  or devices that act as spin filters such as a quantum dot \cite{Rec00} or 
a quantum point contact 
 \cite{Pot02}.

\paragraph{Setup.}

We begin with the description of the setup, shown in Fig. \ref{figset}(a).
Electrons escaping from thermal 
\linebreak
reservoirs with momenta 
\( \mathbf{p}_{1}\simeq -\mathbf{p}_{2} \)
are filtered by two quantum point contacts (QPC) and injected into a
2DEG where they scatter off each other.  To collect after the
collision only  the entangled singlets (EPR pairs), we place two detectors such that
only collisions of electrons 
with final momenta \( \mathbf{p}_{1}',\mathbf{p}_{2}' \) and with scattering
angle \( \theta \in [\pi /2-\delta \theta ,\pi /2+\delta \theta ] \)
are registered. Below we estimate the expected singlet flux (current).

As seen from  Fig. \ref{figset}(b),
the  energies of both incident electrons need to be known, in general, 
to determine the scattering angle
 $\theta \simeq \angle(\mathbf{p}',\mathbf{p})$.
However, for the special case with
opposite momenta $\mathbf{p}_{2}\simeq -\mathbf{p}_{1}$,
$\theta$ is easily determined by
$\theta \simeq \angle(\mathbf{p}_1',\mathbf{p}_{1})$.
Moreover, the energies are then individually conserved,
$ p_{1}\simeq p_{2}\simeq p_{1}'\simeq p_{2}'$ ($ p_{i}=|\mathbf{p}_{i}|$),
which ensures that the outgoing scattering states are unoccupied ($p'_{1,2}>k_F$).
Finally, 
to have well-defined quasiparticle states with long lifetimes,
we assume the electrons to be injected with small excitation energies
$ \xi_i=\hbar ^{2}p_{i}^{2}/2m-E_{F} \ll E_{F}$,
where  $ E_{F}=\hbar ^{2}k_{F}^{2}/2m $ is the Fermi energy 
and $ m$ the effective mass.

\paragraph{Scattering t-matrix.}

 Let us evaluate
the scattering \emph{t-}matrix for two
electrons in the presence of the Fermi sea.
The condition
\( \mathbf{p}_{2}=-\mathbf{p}_{1} \)
defines the Cooper channel, and thus we can follow 
the work by Kohn and Luttinger on  interaction-induced superconductivity 
in a 3D Fermi liquid \cite{Lut66}. 
In contrast to  Ref. \cite{Lut66},
we consider here a 2D system where the screened Coulomb potential
is non-analytic.
Moreover, we need
the complete angular dependence of the scattering amplitude 
---rather than 
only the asymptotics of its Legendre or Fourier coefficients.
The \emph{t-}matrix can be obtained from the (direct) vertex $\Gamma$  
governed by
the Bethe-Salpeter equation \cite{Lut66,FW71}
\begin{eqnarray}
\Gamma (\tilde{p}',\tilde{p};\tilde{P}) & = & \Lambda (\tilde{p}',\tilde{p};\tilde{P})\nonumber \\
 &  & \hspace {-20mm}+\frac{i}{\hbar (2\pi )^{3}}\int d\tilde{k}\Lambda (\tilde{k},\tilde{p};\tilde{P})
 G(\tilde{k}_{1})G(\tilde{k}_{2})\Gamma (\tilde{p}',\tilde{k};\tilde{P}),\label{BS} 
\end{eqnarray}
where $G$ is the single-particle Green function and \( \Lambda  \) the  
irreducible  vertex (we suppress the spin indices as spin is conserved).
The relative and total 4-momenta are denoted by $ \tilde p 
=(\tilde{p}_{1}-\tilde{p}_{2})/2$,
 and $ \tilde P 
=\tilde{p}_{1}+\tilde{p}_{2}$, where \( \tilde{p}_{i}=(\mathbf{p}_{i},\omega _{i}) \)  
with the frequencies \( \omega _{i} \).
The full vertex for singlet/triplets contains the direct and exchange 
parts, i.e. $\Gamma (\tilde{p}',\tilde{p};\tilde{P}) \pm \Gamma 
(-\tilde{p}',\tilde{p};\tilde{P})$.
From the 
vertex we obtain  the \emph{t-}matrix: \( t=\Gamma (\omega _{1}+\omega _{2}\to \xi 
_{1}+\xi_{2}) \) \cite{FW71},
and from it the (unsymmetrized) scattering length \( \lambda =|f|^{2} \)
via the 2D scattering amplitude \( f=-tm/\hbar ^{2}\sqrt{2\pi p} \)
\cite{Bar83}. The scattering lengths for singlet/triplets are then 
given by \( \lambda_{S/T}(\theta) =|f(\theta)\pm f(\theta-\pi)|^{2} \).

The bare Coulomb interaction in 2D is given by 
\( V_{C}(\mathbf{q})
=2\pi e_{0}^{2}/q \)
with \( e_{0}^{2}=e^{2}/4\pi \epsilon _{0}\epsilon _{r} \) and the
dielectric constant \( \epsilon _{r} \). In a first stage, we use
the RPA approximation to account for screening by the Fermi sea \cite{FW71};
this yields a renormalized \( G \) and a screened interaction \( V(\tilde{q})
=V_{C}(q)/[1-V_{C}(q)\chi ^{0}(\tilde{q})] \)
given in terms of the bubble susceptibility diagram \( \chi ^{0} \) with 
\( \tilde{q}=(\mathbf{q},\omega _{q})=\tilde{p}'-\tilde{p} \)
the momentum transfer. We can consider \cite{polarizator} the usual
static Thomas-Fermi screening \( V(q)=e_{0}^{2}/(q+k_{s}) \), with
the screening momentum \( k_{s}=2me_{0}^{2}/\hbar ^{2} \) 
and \( r_{s}=k_{s}/k_{F}\sqrt{2} \).
Note that RPA requires a high density, \( r_{s}\ll 1 \).

In addition to the single interaction line \( V(q) \), the irreducible
vertex \( \Lambda  \) contains, in lowest order in \( V \), two
diagrams: the crossed diagram and the ``wave function modification''
shown in Fig. 2(c) and (d) of Ref. \cite{Lut66}. It is easy to see
that these are smaller than \( V \) by a factor \( r_{s} \)
\cite{Chu89}. Altogether, this justifies the ladder approximation, 
which consists in keeping only the single interaction
line in \( \Lambda \simeq V \). Note that the ladder approximation
requires 
\( \lambda  k_{F}\ll 1 \) \cite{FW71}.
This  condition is consistent with RPA: 
e.g., in the Born approximation \( t\simeq V_{C}\then  \)
\( \lambda k_{F}\sim [V_{C}(k_{F})m]^{2}/2\pi \hbar ^{4}\sim (k_{s}/k_{F})^{2} \).
Finally, we can approximate $G$ by  the free propagator
\( G_{0} \) \cite{Lut66}.
We start with the zero temperature \( T=0 \) case,
and discuss finite \( T \) effects later.

Before solving Eq.(\ref{BS}) to all orders, we first consider its
second-order iteration \( \Gamma ^{(2)}(\mathbf{p}',\mathbf{p};\tilde{P})
=V(\mathbf{p}'-\mathbf{p})+i/\hbar (2\pi )^{3}\, 
\int d\tilde{k}V(\mathbf{k}-\mathbf{p})G(\tilde{k}_{1})G(\tilde{k}_{2})V(\mathbf{p}'-\mathbf{k}) \).
The \( \omega  \)--integration of the Green functions 
yields \cite{FW71}\begin{eqnarray}
\frac{i}{2\pi \hbar }\int d\omega _{k}G_{0}(\mathbf{k}_{1},\frac{\Omega}{2} 
+\omega _{k})G_{0}(\mathbf{k}_{2},\frac{\Omega}{2}
 -\omega _{k}) &  & \nonumber \\
=\frac{N(\mathbf{k}_{1},\mathbf{k}_{2})}
{\hbar\Omega -\xi _{k_{1}}-\xi _{k_{2}}+2i\eta N(\mathbf{k}_{1},
\mathbf{k}_{2})}=:D(\mathbf{k}_{1},\mathbf{k}_{2}), &  & \label{eq2} 
\end{eqnarray}
 with \( N(\mathbf{k}_{1},\mathbf{k}_{2})=1-n(k_{1})-n(k_{2}) \), 
 \( n(k)=\Theta (-\xi _k) \), 
and the \( \tilde{P} \)--frequency is \( \hbar \Omega \to \xi _{1}+\xi _{2} \)
to obtain the \emph{t-}matrix. We now take \( \mathbf{p}=\mathbf{p}_{1}=-\mathbf{p}_{2},\mathbf{k}=
\mathbf{k}_{1}=-\mathbf{k}_{2} \) so that \(  q=|\mathbf{p}'-\mathbf{p}|=2k_{F}|\sin \theta /2| \).
This yields a single
sharp edge in the numerator when \( \xi _{k}=0 \), which coincides
with the zero of the denominator at \( \xi _{k}=\xi  \) 
for incident electrons with vanishing excitation energies
\( \xi =\xi _{1,2}\to 0 \).
Thus, the main contribution to the energy integration
comes from virtual states at the Fermi surface, i.e. \( \xi _{k}\simeq 0 \).
We set \( k=k_{F} \) in \( V \), and integrate only on \( D(k)=D(\mathbf{k},\mathbf{k}) \).
The dominant 
contribution
comes from 
the vicinity of the Fermi sphere \( k=k_{F} \). 
Introducing the notation
\(\nu := (1/2 \pi )\int ^{\infty }_{0}dk\, k\, D(k) \), we 
obtain
\begin{equation}
\label{log-div}
\nu \simeq \frac{m}{2\pi \hbar ^{2}}\log \frac{\xi }{E_{F}}.
\end{equation}
For finite temperatures with \( \xi \ll k_{B}T\ll E_{F} \),
the occupation function \( n(k)=(1+e^{\xi _{k}/k_{B}T})^{-1} \) cuts the 
log divergence 
and $\xi$ is replaced by $ k_{B}T$ in 
Eq. (\ref{log-div}).
This logarithmic divergence  
reveals a 2D Cooper singularity (see below) very 
much like in 3D \cite{FW71}. 
Next, we examine the situation slightly away from the Cooper channel, 
i.e. for non-vanishing total momentum.
Indeed, for experimental reasons it is preferable 
to have a small but 
finite angle \( 2\alpha =\angle (\mathbf{p}_{1},-\mathbf{p}_{2}) \)
between the incident particles to avoid flux misalignment with no 
collision at all. 
In this case, at \( T=0 \) and for \( p_{1}=p_{2} \), we find that 
$ N(k,\phi )	$$
\simeq		$$
\protect{\Theta (k-k_{F}-p\alpha |\sin \phi |)} - 
\protect{\Theta (k_{F}-k-p\alpha |\sin \phi|)} $
depends on the integration angle \( \phi =\angle (\mathbf{k},\mathbf{p}) \)
and, as a consequence, we
find now $\nu (\phi )\simeq (m/2\pi \hbar^{2}) \log \left( 2\alpha |\sin \phi |\right)$ 
in the limit \( p\to k_{F} \).
Thus
the cut-off of
the log-divergence is determined
by \( \max \{\alpha |\sin \phi |,\xi /E_{F},k_{B}T/E_{F}\} \).
However, we recall 
that both outgoing scattering states must be unoccupied, which at $T=0$ requires 
\( p'_{1,2}>k_{F} \).
For \( \pi /2 \)-scattering, \( p'_{1,2}=p_{1}(\cos \alpha \pm \sin \alpha )\then  \)
\( p_{1}\agt k_{F}(1+\alpha ) \); see Fig. \ref{figset}(b).
Therefore, $\alpha <\xi/E_{F}$
and the relevant cut-off is still given by $\xi$ or $k_B T$.

We repeat this procedure in each order in \( V \) and rewrite
Eq. (\ref{BS}) as
$ t(\theta) $$
=           $$
v(\theta )  $$
+	    $$
(\nu /2\pi )\int \mathrm{d}\phi v(\phi )t(\theta -\phi )$,
 with 
$ v(\phi )  $$
=           $$ 
2\pi e^{2}/(2k_{F}|\sin \phi /2|+k_{s})$
the 2D Coulomb potential V at the Fermi 
surface.
To solve this equation, we expand \( v \) and \( t \) into Fourier series: 
$v(\phi )$$
=        $$
\sum_{n}v_{n}e^{in\phi }$, etc. 
We finally get for the {\emph t-}matrix at the Fermi surface 
\begin{equation}
\label{tmatrix result}
t(\theta )=\sum _{n}\frac{v_{n}}{1-\nu v_{n}}e^{in\theta },
\end{equation}
with the Fourier coefficients 
\begin{equation}
\label{vn}
v_{n}=\frac{4e_{0}^{2}}{k_{F}\cos \gamma }\sum _{\mathrm{odd}\, m\geq 
1}\frac{\cos (m\gamma )}{2n+m}
\end{equation}
and \( \gamma =\mathrm{arcsin}\left( r_{s}/\sqrt{2}\right)  \).
Below
we use this result to evaluate the scattering length. 

\paragraph{Scattering length. }

To illustrate 
how the scattering process gets renormalized by the Fermi sea, 
we compare 
the above result (\ref{tmatrix result}) with  the {\em t}-matrix 
 \( |t_{\mathrm{C}}(\theta )|
=e_{0}\hbar \left[ \tanh \left( \pi me_{0}^{2}/k_{F}\hbar ^{2}\right) 
\pi /mk_{F}\sin ^{2}(\theta /2)\right] ^{1/2} \)  
obtained for the bare Coulomb potential \( V_{C} \) in 2D \cite{Bar83}.
We 
 use typical parameters for a GaAs  2DEG: \( \epsilon _{r}=13.1 \),
\( r_{s}=0.86 \) and a sheet density \( n=4 \cdot 10^{15}\, \mathrm{m}^{-2} \)
\cite{And82}, and assume
 \( \xi < k_{B}T=10^{-2}E_{F} \) (\( T=2 \) K).
\begin{figure}[!tb]
{\centering \resizebox*{1\columnwidth}{!}
{\includegraphics{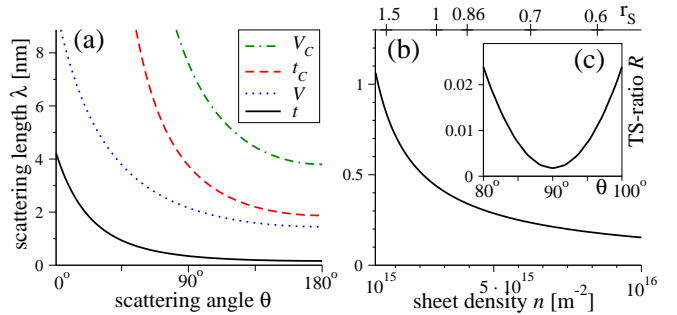}} \par}
\caption{ {\small \label{figres} 
Plots of scattering quantities obtained from the \emph{t-}matrix (\ref{tmatrix result})
for \protect\( r_{s}=0.86\protect \) (GaAs) and \protect\( k_{B}T/E_{F}=10^{-2}\protect \).
(a) Angular dependence of the scattering length 
\protect\( \lambda (\theta )\protect \).
We compare \protect\( \lambda \protect \) to its  Born
approximation given by \protect\( t\simeq V\protect \) and to the
bare scattering of two particles (no Fermi sea) given either by the exact
result (\protect\( t_{C}\protect \)) or by the first-order (\protect\( V_{C}\protect \)).
(b) Dependence of \protect\( \lambda (\pi /2)\protect \) on the sheet
density \protect\( n\protect \) (see the corresponding 
\protect\( r_{s}=me_{0}^{2}/\hbar ^{2}\sqrt{\pi n}\protect \)
in the top axis). 
(c) TS-ratio \protect\( 
R(\theta,\delta\theta)\protect \)
of triplets/singlets  
detected at a scattering angle \protect\( \theta \protect \),
for an aperture \protect\( \delta \theta =5^{\mathrm{o}}\protect \).\vspace{-5mm}}
}
\end{figure}
In Fig. \ref{figres}(a) we plot the scattering length \( \lambda (\theta ) =|tm/\hbar ^{2}\sqrt{2\pi p}|^{2} \)
as function of the scattering angle $\theta$
(without antisymmetrization).  The reduction in amplitude due to the
Fermi sea is seen to be quite substantial (compared to the bare \( t_{C} \))  which can be traced back
to the relatively large screening \( k_{s} \) (\( r_{s}=0.86 \)).
We also have \( t\to 0 \) as \( k_{\mathrm{B}}T,\xi \to 0 \).
For very small \( r_{s} \) or increasing $T$
we can drop \( \nu v_{n}\ll 1 \) in Eq. (\ref{tmatrix result}),
and 
thus 
recover the Born approximation  with the bare 
potential
\( V_{C} \).

In addition, the higher-order terms appearing in the iteration of
the Bethe-Salpeter equation further reduce the scattering (compare
\( t \) to the first order \( V \)) \cite{borncooper}.  In Fig. \ref{figres}(b)
we see the significant reduction of the scattering length $\lambda$ as the density $n$
is increased, which could be tested experimentally.

\paragraph{Production of EPR-pairs.}

We now turn to the production of entangled electrons in the spin-singlet
state. We consider detectors placed at \( \theta  \), with
a small aperture angle of \( 2\delta \theta  \) (Fig. \ref{figset}).
We introduce the scattering lengths around $\theta$ (integrated over \( 2\delta \theta \)) 
for singlets ($+$) and triplets ($-$):
\( \bar{\lambda}_{S/T}(\theta)=
2\int ^{\theta}_{\theta-\delta \theta }d\theta' |f(\theta' )\pm f(\pi -\theta' |^{2} \).
A useful measure is the ratio between the number 
$N_{T/S}$ of 
detected triplets/singlet, 
$R(\theta,\delta\theta)=N_{T}/N_{S}={3\bar{\lambda}_{T}}/{\bar{\lambda}_{S}}$.
Next, expanding for small $\delta \theta$ around $\theta=\pi/2$ we find in leading order
\( \bar{\lambda}_{S}\simeq 8|f(\pi /2)|^{2}\delta \theta\),  
and \( \bar{\lambda}_{T}\simeq (8/3)|f'(\pi /2)|^{2}\delta \theta ^{3} \), 
which yields
\begin{equation}
\label{i}
R(\pi/2,\delta\theta)\simeq \left|
\frac{f'(\pi /2)}{f(\pi /2)}\right| ^{2}\delta \theta ^{2} \, .
\label{ratio} 
\end{equation}
 Using Eq. (\ref{tmatrix result}) we find \( |f'/f|=0.48 \) at $\theta =\pi/2$
for GaAs, and note that this ratio remains of order unity for a wide
parameter range \( k_{B}T/E_{F}=10^{-1}-10^{-10} \), and \( r_{s}=0.1-1 \).
The Born approximation \( |V'/V|=1/2(1+r_{s})\simeq 0.266 \) is approached
for $k_B T/E_F > 10^{-1}$. 
Therefore, this setup 
indeed
allows the production of dominantly singlets (EPR pairs) at the
detectors 1 and 2 provided that
the aperture angle is sufficiently small. 
For example, \( R(90^{\mathrm{o}},5^{\mathrm{o}})=0.2\% \)
or \( R(85^{\mathrm{o}},5^{\mathrm{o}})=0.7\% \); 
see Fig. \ref{figres}(c).

To estimate the singlet current for a given input current $I$
we assume that the incident electrons occupy the lowest transverse mode
in the QPC, giving plane waves of transverse width $w$ (typically \( w\simeq 100\, \mathrm{nm} \)). 
The probability for the singlets to be scattered into the detectors is
\( P_{S}=(1/4)\bar {\lambda}_{S}/w=0.06 \% \) with
\( \bar{\lambda} _{S}= 0.24\, \mathrm{nm} \) for \( \delta \theta =5^{\mathrm{o}} \).
It is advantageous to inject simultaneously the two electrons from 
the reservoirs (e.g., by opening both QPCs 
 simultaneously) \cite{nonsim}.
Then the singlet current is given by \( I_{S}=P_{S}I = 0.6\, \mathrm{pA} \) for 
\( I = 1\, \mathrm{nA} \).
The total scattering length for unpolarized 
electrons is 
\( \lambda _{\mathrm{tot}}=(1/4)\int _{-\pi /2}^{\pi /2}d\theta [\lambda _{S}(\theta )
+3\lambda _{T}(\theta )] =3.4 \)  nm in GaAs (compared to \( 11 \) nm in Born approximation). 
This is consistent with the
ladder approximation (\( \lambda _{\mathrm{tot}}k_{F}=0.53 \)), and 
yields  the total scattering probability
\( P_{\mathrm{tot}}=\lambda _{\mathrm{tot}}/w = 3 \% \).

An important requirement is  that only correlated 
electrons (singlets) are 
counted at the detectors 1 and 2, i.e. electrons which have 
scattered off
each other at \( \theta =\pi /2 \), whereas we need to avoid the counting 
of uncorrelated electrons which are accidentally scattered into  the 
detectors due e.g. to  impurity scattering. This requirement could be 
fulfilled e.g.
by coincidence measurements and/or with the help of an ac 
modulation applied to each reservoir
with different frequencies \( \omega _{1} \) and \( \omega _{2} \).
This would enable a frequency selection of the electrons, since only 
the electrons which have interacted
are modulated by the two frequencies \( \omega _{1}\pm \omega _{2} \).

Besides the  observation of the  scattering length
and its density and angular dependence, 
our proposal could be further tested by adding a beam-splitter to probe the
singlet-state via noise measurement \cite{Bur00,noisereview}, 
by performing tests of Bell's inequality \cite{Kaw01,noisereview}
with spin filters \cite{Rec00,Pot02}, or
by using p-i-n junctions \cite{Fie99} to transform
singlets into entangled photon pairs.
An alternative test requires
spin-filters, obtained e.g. 
by tuning the QPCs into the spin-filtering regime \cite{Pot02},
or by replacing them by spin-filtering quantum dots \cite{Rec00}.
Then with increasing spin polarization \( \mathcal{P}\) 
the probability of incoming singlets \( \rho _{S}=(1-\mathcal{P}^{2})/2 \)
is suppressed, and 
the singlet current at \( \theta =\pi /2 \) vanishes
at full polarization \( \mathcal{P} =1\).

We now comment on the Kohn-Luttinger instability
\cite{Lut66}. 
The crossed diagram 
---which in 2D can lead to an instability only for excitations 
with $q>2 k_F$ \cite{Chu93}--- does not
lead to a strong renormalization of the scattering vertex, as
the associated temperature is infinitesimal, $k_B T/E_F \sim {\rm exp}(-10^3)$ \cite{SAL}.
This is larger than the value $\sim {\rm exp}(-10^5)$ found in 3D \cite{Lut66}, despite
the asymptotic decay for \( v_{n}\sim n^{-2} \)  being slower than
in 3D, \( v_{l}\sim e^{-l} \)
 (the decay is polynomial because the 2D potential is non-analytic). 
The crossed diagram, given by the polarization propagator, 
has asymptotics \( \sim n^{-3/2} \), instead of \( \sim l^{-4} \) in 3D. 
Finally, we also checked
that the repulsive electron-electron interaction  is not appreciably affected 
by polar or acoustic phonons \cite{SAL}.

In conclusion, we have shown that the scattering process of two electrons
 can be used
to isolate and select the entangled spin-singlets at the scattering angle $\pi/2$.
The process survives in the presence of an interacting Fermi sea 
which, however,
reduces the scattering amplitude substantially.
Nevertheless, according to our analysis
the current of entangled electrons is within experimental reach
for realistic two-dimensional electron gases.

\begin{acknowledgments}
{\footnotesize \vspace{-0mm}}We thank C. Egues, V. Golovach and
W. Coish for useful discussions. This work has been supported by NCCR
{}``Nanoscale Science'', Swiss NSF, DARPA and ARO.{\footnotesize \vspace{-5mm}} 
\end{acknowledgments}


\begin{thebibliography}{10}
\bibitem{borncooper}
Outside the Cooper channel, in the absence of the logarithmic divergence 
(3), the Born approximation is valid.
\bibitem{nonsim}
For random injection times, the singlet current would scale down with the 
probability occupation for $\ket{\mathbf{p}_{1} \mathbf{p}_{2}}$,
given roughly by 
$P_{12}\sim (e W_{\rm out}/I)^2$, with $W_{\rm out}$ the rate for electrons to escape out
of the scattering region.
\end{thebibliography}
\end{document}